\documentclass{Interspeech}

\interspeechcameraready

\usepackage[utf8]{inputenc} % allow utf-8 input
\usepackage[T1]{fontenc}    % use 8-bit T1 fonts
\usepackage{hyperref}       % hyperlinks
\usepackage{url}            % simple URL typesetting
\usepackage{booktabs}       % professional-quality tables
\usepackage{amsfonts}       % blackboard math symbols
\usepackage{nicefrac}       % compact symbols for 1/2, etc.
\usepackage{microtype}      % microtypography
\usepackage{xcolor}         % colors
\usepackage{multirow}
\usepackage{graphicx}
\usepackage{array}    % For advanced table handling

\newcounter{rownum}   % Define a counter
\newcommand{\rownumber}{\stepcounter{rownum}\arabic{rownum}} % Create command to automatically increment the counter

\PassOptionsToPackage{hyphens}{url}
\newcommand\blfootnote[1]{%
  \begingroup
  \renewcommand\thefootnote{}\footnote{#1}%
  \addtocounter{footnote}{-1}%
  \endgroup
}

\title{Audiobox TTA-RAG: Improving Zero-Shot and Few-Shot Text-To-Audio \\ with Retrieval-Augmented Generation}

\author[affiliation={1, \dagger}]{Mu}{Yang}
\author[affiliation={2}]{Bowen}{Shi}
\author[affiliation={2}]{Matthew}{Le}
\author[affiliation={2}]{Wei-Ning}{Hsu}
\author[affiliation={2}]{Andros}{Tjandra}
\affiliation{Center for Robust Speech Systems (CRSS)}{University of Texas at Dallas}{USA}
\affiliation{}{Meta AI}{USA}
\email{mu.yang@utdallas.edu, androstj@meta.com}


\begin{document}

\maketitle
\blfootnote{$^\dagger$Work done during an internship at Meta.}
\begin{abstract}
This work focuses on improving Text-To-Audio (TTA) generation on zero-shot and few-shot settings (i.e. generating unseen or uncommon audio events).
Inspired by the success of Retrieval-Augmented Generation (RAG) in Large Language Models, we propose \emph{Audiobox TTA-RAG}, a novel retrieval-augmented TTA approach based on Audiobox, a flow-matching audio generation model.
Unlike the vanilla Audiobox TTA solution that generates audio conditioned on text only, we extend the TTA process by augmenting the conditioning input with both text and retrieved audio samples. 
Our retrieval method does not require the external database to have labeled audio, offering more practical use cases. 
We show that the proposed model can effectively leverage the retrieved audio samples and significantly improve zero-shot and few-shot TTA performance, with large margins on multiple evaluation metrics, while maintaining the ability to generate semantically aligned audio for the in-domain setting.\footnote{Audio samples are available at \url{https://tta-rag-is25.github.io/}}

\end{abstract}

\section{Introduction}
Text-to-audio (TTA) synthesis is a task of generating audio given a textual description (e.g. \textit{bird chirping}, \textit{car honking}, etc.). Recently, generative models have significantly advanced the field. These include diffusion-based \cite{vyas2023audiobox, ghosal2023text, yang2023diffsound, liu2023audioldm, liu2024audioldm, huang2023make, saito2024soundctm, zhu2025cosyaudio} and autoregressive audio language model-based methods \cite{kreuk2023audiogen, yang2023uniaudio, copet2023simple}.

One limitation of the current leading TTA methods is the generalization ability to few-shot and zero-shot scenarios. For few-shot TTA, as pointed out by \cite{yuan2024retrieval}, TTA models suffer from the \emph{long-tailed generation} problem: due to data scarcity, the TTA models struggle to learn the text-to-audio distribution mapping for audio events that appear only a few times in the training set. For the more challenging zero-shot TTA, i.e. generating unseen audio events, the performance further degrades. 

To improve few-shot and zero-shot TTA performance, we extend the TTA process with Retrieval-Augmented Generation (RAG). RAG, originally proposed in Large Language Models (LLM) studies, has emerged as a promising approach to guiding LLMs to generate accurate, grounded responses \cite{guu2020retrieval, lewis2020retrieval}. 
It augments the input prompts by retrieving relevant information from external databases and provides the LLMs with more informative contexts, enhancing the performance on knowledge-intensive tasks \cite{guu2020retrieval, lewis2020retrieval, asai2022evidentiality, glass2021robust}.
For zero-shot and few-shot TTA tasks, we hypothesize that RAG can also benefit TTA by extending the conditioning input with retrieved relevant audio, which serve as supplementary ``exemplars'' contexts to guide the generation.

We propose \emph{Audiobox TTA-RAG}, a novel retrieval-augmented TTA approach based on Audiobox \cite{vyas2023audiobox}, a state-of-the-art audio generation model. In addition to the text description, the inputs also include audio samples retrieved from an external database. Specifically, the retrieved audio samples are processed by a novel text-conditioned retrieval audio encoder which extracts acoustic information using the text description as the query. The final conditioning input for TTA is formed by concatenating the encoder output with the text embedding. Unlike the vanilla TTA model where conditioning inputs only contain textual modality, the proposed TTA-RAG model provides both textual and acoustic modality as the conditioning context, offering rich and diverse grounding information. 
During training, we retrieve audio samples by performing audio-to-audio (A2A) matching to a large audio database using the target audio as query. 
During inference, since the target audio is unavailable, we perform text-to-audio (T2A) matching using the text description as query. Note that our approach does not require the external database to have \emph{labeled} audio, thus unlocking the potential to retrieve ``from wild'' and offering more practical use cases. 
To evaluate the proposed model, we curated zero-shot and few-shot TTA test sets containing unseen or uncommon audio events in the training data. 
Our empirical results show that, training Audiobox TTA-RAG on AudioCaps \cite{kim2019audiocaps}, with AudioSet \cite{gemmeke2017audio} as the retrieval source, we can significantly improve zero-shot and few-shot TTA performance, with large margins on multiple evaluation metrics. To summarize, this work makes the following contributions:

\begin{itemize}
    \item To evaluate zero-shot and few-shot TTA performance, we curated evaluation datasets that contain audio events that are unseen or uncommon in the training set.
    \item We design a novel TTA-RAG model that can effectively retrieve unlabeled audio and significantly improve zero-shot and few-shot TTA performance.
    \item We retrieve audio using different retrieval methods and different source databases during training and inference, and analyze their effects on the TTA performance.
\end{itemize}

\begin{figure*}[t]
    \centering
    % \captionsetup{skip=5pt}
    \includegraphics[width=\textwidth]{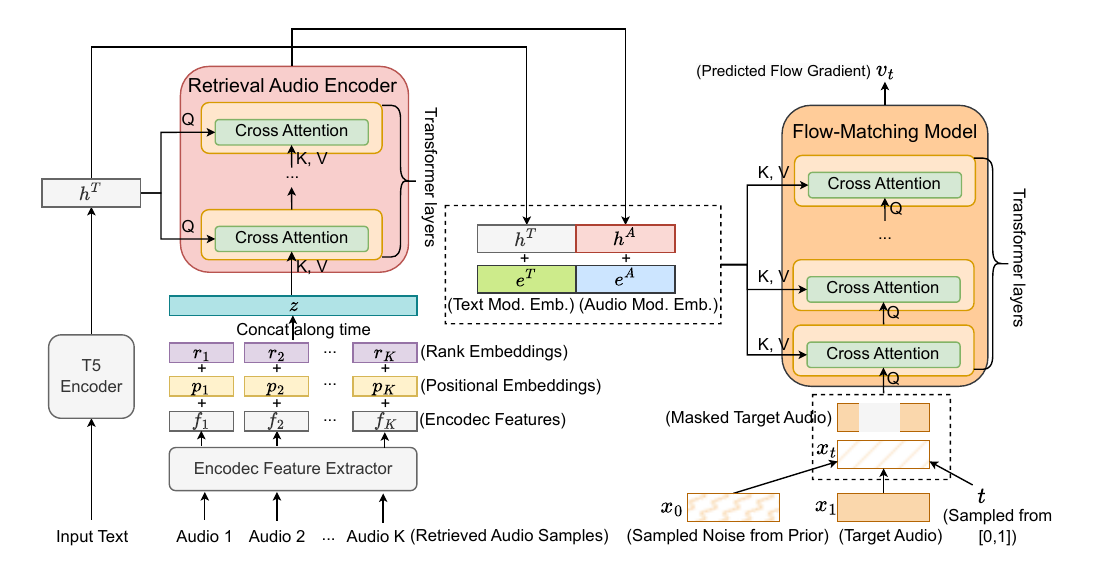}
    \vspace{-7mm}
    \caption{The proposed Audiobox TTA-RAG framework. Grey blocks indicate pre-trained and frozen modules. The remaining modules are trained from scratch in an end-to-end manner.}
    \vspace{-3mm}
    \label{fig:system}
\end{figure*}
\section{Related work}

Following the success of RAG on LLM-based text generation, RAG has also been extensively studied for text-to-image generation \cite{blattmann2022retrieval, sheynin2023knndiffusion} or text-image multi-modal generation \cite{yasunaga2023retrieval}. 
For audio tasks, only a limited number of prior works have investigated RAG. 
Xue et al. \cite{xue24b_interspeech} applied a retrieval method to select text and audio prompts to enhance a prompt-based text-to-speech system. 
RECAP \cite{ghosh2024recap} focused on audio captioning. They augmented the conditioning input to an audio-captioning model with top-k captions from an external datastore using the audio as query and showed that such a retrieval-augmented captioning model has better performance for out-of-domain settings. 
\cite{min2024speech} approached speech-based Question Answering using a speech retriever, which retrieves relevant speech passages given a speech query. Conditioned on the retrieved audio passages, a speech-language model generates text responses, without the need to use a separate speech recognition model to transcribe the speech query. 
\cite{ogura2024retrieval} proposed a retrieval-augmented approach for Anomalous Sound Detection, where the text caption of an anomalous sound, and k text captions of normal sounds are used to prompt GPT-4 to describe the difference between the anomalous sound and normal sounds.

Our work is most similar to Re-AudioLDM \cite{yuan2024retrieval}, which leveraged both retrieved audio and the corresponding captions for TTA generation. 
Unlike Re-AudioLDM, our approach does not use the corresponding captions of the retrieved audio as the conditional inputs. The proposed TTA-RAG method does not require labeled retrieval sources, enabling us to query larger-scale unlabeled databases for more diverse samples.

\section{Method}

\subsection{Curating Zero-shot and Few-shot Evaluation Set}
Our model is trained on AudioCaps \cite{kim2019audiocaps} dataset, which contains a subset of audio in AudioSet \cite{gemmeke2017audio} with human-annotated captions. Since our main goal is to evaluate zero-shot and few-shot TTA performance, i.e. generating audio events that are unseen or uncommon in AudioCaps, we curate the zero-shot and few-shot evaluations set from AudioSet. 
AudioSet provides short text tags in a hierarchical ontology from abstract to specific. For example, an audio with a \emph{coo} sound event may have the following tags: \emph{Animal $\rightarrow$ Wild animals $\rightarrow$ Pigeon,dove $\rightarrow$ Coo}. We go through AudioSet audio and select those with leaf tags (most specific audio events) that do not appear (zero-shot) or appear less than 3 times (few-shot) in AudioCaps. We ensure that the selected audio samples are not included in AudioCaps. Finally, we adopt Audio-Flamingo-generated \cite{kong2024audio} captions as input texts for our TTA evaluation. We end up with 5546 and 601 samples for zero-shot and few-shot set, respectively.

\subsection{Retrieval method}
Our goal is to retrieve audio samples as the additional context for the TTA process. During training, we perform audio-to-audio (A2A) retrieval. Specifically, we compute CLAP \cite{wu2023large} audio embeddings of the target audio and the audio in the source database.\footnote{We use the \emph{630k-audioset-fusion-best} checkpoint in \url{https://github.com/LAION-AI/CLAP}.} Then we keep the top-k audio samples with the highest embedding cosine similarities to the target audio. During inference, since the target audio is not available, we perform text-to-audio (T2A) retrieval, i.e. computing the cosine similarity between the input text embedding (given by the CLAP text encoder) and source audio embeddings (given by the CLAP audio encoder). We investigated using T2A retrieval during training and found that the TTA performance is inferior to using A2A retrieval (see our analyses in Section \ref{sec:res-zero}). To avoid information leakage, we ensure that the target audio itself is excluded from the retrieval source.

\begin{table*}[t]
\caption{TTA performance on zero-shot and few-shot evaluation sets. \emph{Retrieval Number} denotes the number of retrieved audio samples. \emph{Retrieval Setting} describes the retrieval method (T2A or A2A matching) and retrieval data source (AC - AudioCaps; AS - AudioSet) used in training and inference.}
\label{tab-zero-shot}
\resizebox{\textwidth}{!}{
\def\arraystretch{1.2}
\begin{tabular}{>{\centering\arraybackslash}p{0.5cm}ccccccccc}
\hline
ID & \begin{tabular}[c]{@{}c@{}}Evaluation \\ Dataset\end{tabular} & Model                             & \begin{tabular}[c]{@{}c@{}}Retrieval \\ Number\end{tabular} & \begin{tabular}[c]{@{}c@{}}Retrieval Setting\\ (Training)\end{tabular} & \begin{tabular}[c]{@{}c@{}}Retrieval Setting\\ (Inference)\end{tabular} & IS $\uparrow$            & CLAP (\%) $\uparrow$     & KL $\downarrow$           & FAD $\downarrow$          \\ \hline

\rownumber & \multirow{6}{*}{Zero-shot}                                    & Audiobox TTA                      & N/A                                                         & N/A                                                                    & N/A                                                                     & 4.95          & 20.40          & 3.83          & 9.31          \\ \cline{3-10} 
\rownumber &                                                              & \multirow{5}{*}{Audiobox TTA-RAG} & 3                                                           & T2A - AC                                                               & T2A - AS                                                                & 5.32          & 20.92          & 3.80          & 8.08          \\
\rownumber &                                                              &                                   & 3                                                           & A2A - AC                                                               & T2A - AS                                                                & 5.33          & 24.90          & 3.49          & 8.11          \\
\rownumber &                                                              &                                   & 3                                                           & T2A - AS                                                               & T2A - AS                                                                & 5.45          & 21.19          & 3.74          & 8.68          \\
\rownumber &                                                              &                                   & 3                                                           & A2A - AS                                                               & T2A - AS                                                                & \textbf{5.63} & 27.71          & \textbf{3.02} & \textbf{7.42} \\
\rownumber &                                                              &                                   & 10                                                          & A2A - AS                                                               & T2A - AS                                                                & 5.58          & \textbf{29.01} & 3.08          & 8.07          \\ \hline \hline 
\rownumber & \multirow{6}{*}{Few-shot}                                     & Audiobox TTA                      & N/A                                                         & N/A                                                                    & N/A                                                                     & 5.87          & 23.98          & 4.37          & 2.70          \\ \cline{3-10} 
\rownumber &                                                              & \multirow{5}{*}{Audiobox TTA-RAG} & 3                                                           & T2A - AC                                                               & T2A - AS                                                                & 6.00          & 23.75          & 4.51          & 2.94          \\
\rownumber &                                                              &                                   & 3                                                           & A2A - AC                                                               & T2A - AS                                                                & 5.90          & 28.64          & 3.91          & 2.36         \\
\rownumber &                                                              &                                   & 3                                                           & T2A - AS                                                               & T2A - AS                                                                & 5.92          & 24.79          & 4.22          & 2.85          \\
\rownumber &                                                              &                                   & 3                                                           & A2A - AS                                                               & T2A - AS                                                                & 6.07          & 30.12          & 3.94          & \textbf{2.31} \\
\rownumber &                                                              &                                   & 10                                                          & A2A - AS                                                               & T2A - AS                                                                & \textbf{6.61} & \textbf{31.92} & \textbf{3.63} & 2.69          \\ \hline
\end{tabular}
}
\vspace{-2mm}
\end{table*}
\subsection{The Audiobox TTA-RAG Model}
The structure of the proposed Audiobox TTA-RAG is shown in Figure \ref{fig:system}. 
Following \cite{vyas2023audiobox}, we feed the $K$ retrieved audio samples to a pre-trained Encodec \cite{defossez2022high} encoder and extract the pre-quantization features $f_1, f_2, ..., f_K$ where $f_i \in \mathbb{R}^{T_i \times D}$, $T_i$ denotes the time dimension of the $i$-th audio sample and $D$ denotes the feature dimension. A positional embedding \cite{vaswani2017attention} $p_i \in \mathbb{R}^{T_i \times D}$ is added individually to $f_i$ to encode the temporal information within each retrieved audio. Then, a rank embedding $r_i \in \mathbb{R}^D$ (broadcasted along the time dimension) is further added to $f_i$. The rank embeddings come from a learnable embedding matrix with $K$ vectors, each denoting the rank information of an audio among all used retrieval samples. The final input to the Retrieval Audio Encoder $z \in \mathbb{R}^{\left(\sum_{i=1}^{K}T_i\right) \times D}$ is constructed by concatenating $\{f_i+p_i+r_i\}_{i=1}^{K}$ along the time dimension. 

The Retrieval Audio Encoder takes two inputs, $z$ and text embedding $h^T$ from a pre-trained T5-base encoder \cite{raffel2020exploring}, where $h^T$ is used as queries to the cross attention layers to extract useful information from $z$. 
The output of the Retrieval Audio Encoder $h^A$ is concatenated with $h^T$ along time dimension. Two learnable modality embeddings $e^T$ and $e^A$ (both broadcasted along time dimension) are added to $h^T$ and $h^A$, respectively, forming the conditioning input to the Flow-Matching model. 
Compared to the original Audiobox TTA model \cite{vyas2023audiobox}, the additional $h^A$ contains acoustic information extracted from the retrieved audio samples conditioned on the input text. This allows the TTA-RAG model to effectively utilize the supplementary information to generate the target audio.

The whole system is trained end-to-end, following the same Flow-Matching (FM) objective \cite{lipman2023flow} in Audiobox \cite{vyas2023audiobox}. Given a data sample $x_1$ from the data distribution and a flow step $t$ drawn from $\mathcal{U}[0, 1]$, a noisy version $x_t$ and its derivative $v_t = dx_t / dt$ for a chosen conditional path is constructed. The FM model predicts $v_t$ given $x_t$ and $t$. During inference, we start from a sample $x_0$ drawn from a prior distribution. An ordinary differential equation (ODE) solver estimates $x_1$ using $x_0$ and the derivatives given by the FM model.

Specifically, let $x_1$ denote the Encodec pre-quantization feature of an audio sample, and $t \sim \mathcal{U}[0, 1]$ denote a random uniformly sampled time step, 
with the Optimal Transport conditional path we construct
\begin{equation}
    x_t = (1-(1-\sigma_{min})t)x_0 + tx_1
\end{equation}
and
\begin{equation}
    v_t = x_1-(1-\sigma_{min})x_0
\end{equation} 
where $x_0$ is drawn from the prior distribution $N(0, I)$ and $\sigma_{min}$ is a small value ($10^{-5}$). 
The Flow-Matching model $u$ minimizes
\begin{equation}
    \mathbb{E}_{t, x_1, x_0}||u(x_t, t)-v_t||^2
\end{equation}
During training, $x_t$ is concatenated with the partially or fully masked Encodec feature $x_1$. During inference, $x_1$ is set to a fully masked (blank) feature.

\section{Experiments}

\setcounter{rownum}{0}
\begin{table*}[t]
\caption{TTA performance on AudioCaps test set. We also compare to the reported results from a RAG-based prior work \cite{yuan2024retrieval}.}
\label{tab-ac-test}
\resizebox{\textwidth}{!}{
\def\arraystretch{1.2}
\begin{tabular}{>{\centering\arraybackslash}p{0.5cm}cccccccc}
\hline
ID & Model                             & \begin{tabular}[c]{@{}c@{}}Retrieval \\ Number\end{tabular} & \begin{tabular}[c]{@{}c@{}}Retrieval Setting\\ (Training)\end{tabular} & \begin{tabular}[c]{@{}c@{}}Retrieval Setting\\ (Inference)\end{tabular} & IS $\uparrow$           & CLAP (\%) $\uparrow$     & KL $\downarrow$          & FAD $\downarrow$           \\ \hline
\rownumber & AudioLDM-S \cite{yuan2024retrieval}                       & N/A                                                         & N/A                                                                    & N/A                                                                     & 6.48          & 26.75          & 1.63          & 2.31          \\ \hline
\rownumber & Re-AudioLDM-S \cite{yuan2024retrieval}                       & 3                                                           & T2T - AC (A+T)                                                              & T2T - AC (A+T)                                                               & 7.31          & 37.07          & 1.27          & 1.48          \\
\rownumber & Re-AudioLDM-S \cite{yuan2024retrieval}                       & 10                                                          & T2T - AC (A+T)                                                         & T2T - AC (A+T)                                                          & 7.33          & 37.15          & \textbf{1.23} & \textbf{1.40} \\ \hline \hline
\rownumber & Audiobox TTA                      & N/A                                                         & N/A                                                                    & N/A                                                                     & 9.34 & 34.21          & 1.31          & 2.54          \\ \hline
\rownumber & \multirow{7}{*}{Audiobox TTA-RAG} & 3                                                           & T2T - AC (A+T)                                                              & T2T - AC (A+T)                                                               & \textbf{9.41}          & 35.06          & 1.37          & \textbf{2.31}          \\
\rownumber &                                  & 3                                                           & A2A - AS                                                               & T2T - AC                                                                & 8.72          & 34.61          & \textbf{1.29} & 2.37 \\
\rownumber &                                  & 3                                                           & A2A - AS                                                               & T2A - AS                                                                & 8.05          & 37.73          & 1.63          & 4.73          \\
\rownumber &                                  & 3                                                           & A2A - AS                                                               & T2A - AC                                                                & 8.40          & 37.37          & 1.44          & 3.54          \\
\rownumber &                                  & 10                                                          & A2A - AS                                                               & T2T - AC                                                                & 9.16          & 34.71          & 1.33          & 3.00          \\
\rownumber &                                  & 10                                                          & A2A - AS                                                               & T2A - AS                                                                & 8.21          & 37.76          & 1.83          & 6.57          \\
\rownumber &                                  & 10                                                          & A2A - AS                                                               & T2A - AC                                                                & 8.92          & \textbf{38.13} & 1.53          & 4.70          \\ \hline
\end{tabular}
}
% \vspace{-5mm}
\end{table*}

\subsection{Experimental Setup}

\textbf{Datasets} We train our models on the AudioCaps dataset \cite{kim2019audiocaps}, following the official train/dev/test splits. The training set contains 45k+ \textasciitilde10-second audio samples and corresponding human-annotated captions. 
In addition, we also use AudioSet \cite{gemmeke2017audio} as the retrieval source, which contains 1.8M+ \textasciitilde10-second audio samples.\footnote{The dataset used in this study was filtered to comply with several rules and restrictions.} Our models are trained on the AudioCaps training set, and evaluated on the AudioCaps test set, as well as our curated zero-shot and few-shot evaluation sets.

\noindent\textbf{Implementation Details} The retrieval audio encoder consists of 3 Transformer layers, with 16 heads, model dimension 1024 and feed-forward dimension 4096. Other configurations follow \cite{vyas2023audiobox}.  We train the Audiobox TTA baseline on 16 GPUs, with a batch size of 15k tokens per GPU. Our proposed Audiobox TTA-RAG models use 3 or 10 retrieval audio samples, trained on 32 GPUs with a batch size of 8k and 5k tokens, respectively. All models are trained for 150k steps with 5k warm-up steps. We use Adam optimizer with a polynomial decay scheduler and a peak learning rate of 1e-4.

\noindent\textbf{Evaluation Metrics} Following \cite{yuan2024retrieval}, we consider four evaluation metrics: Inception Score (IS), CLAP Score, Kullback-Leibler (KL) Divergence, and Frechet Audio Distance (FAD). IS is a reference-free measure that assigns higher scores for audio with higher variety. CLAP score computes cosine similarity between the output audio embedding and input text embedding, measuring the semantic alignment between audio and text. KL is an instance-level metric computing the divergence of the acoustic event posterior between the reference and the output. FAD measures the distribution-level similarity between reference samples and generated samples. 

\subsection{Performance on Zero-shot and Few-shot Evaluation Set}
\label{sec:res-zero}

Table \ref{tab-zero-shot} shows zero-shot (row 1-6) and few-shot (row 7-12) TTA performance for different model configurations. Zero-shot and few-shot settings focus on audio events that are unseen or uncommon in the AudioCaps training set, thus we only consider retrieving from AudioSet during inference. Since AudioSet does not have human-annotated detailed captions, we perform text-to-audio retrieval (T2A - AS). We investigate different retrieval methods and data sources used in training.

\noindent\textbf{Impact of retrieval method}  Comparing T2A retrieval to A2A retrieval from the same data source (e.g. row 4 vs. row 5, row 10 vs. row 11), we see that in general A2A retrieval outperforms T2A retrieval. This indicates that the retrieved audio used in training plays a major role in guiding the TTA-RAG model to learn how to effectively utilize the additional retrieval context. Our proposed TTA-RAG approach relies on a cross-attention-based retrieval audio encoder that aims to encode useful acoustic information conditioned on text. Intuitively, A2A matching retrieves audio samples that are more acoustically similar to the target audio. Hence, they may offer stronger supervision signals to the retrieval audio encoder, and thus may benefit its training.

\noindent\textbf{Impact of retrieval source and retrieval number} From Table \ref{tab-zero-shot}, we also observe that training on samples retrieved from AudioSet outperforms retrieving from AudioCaps (e.g. row 5 vs. row 3, row 11 vs. row 9), demonstrating the benefit of more diverse retrieved samples. Increasing the retrieval number from 3 to 10 provides further improvements, especially for CLAP score. With RAG, we can significantly improve the TTA basleline by large margins on all metrics.

\subsection{Performance on AudioCaps Test Set}

Table \ref{tab-ac-test} shows the TTA performance on the AudioCaps test set. 
We also include Re-AudioLDM \cite{yuan2024retrieval}, a RAG-based TTA model into the comparison (rows 1-3). 
Re-AudioLDM performs text-to-text (T2T) retrieval from AudioCaps and uses both the retrieval audio and the corresponding text captions as the conditioning inputs (denoted as \emph{T2T - AC (A+T)}). Following the retrieval setting of Re-AudioLDM, we train an Audiobox TTA-RAG model with both retrieved audio and text (row 5), where we concatenate the Encodec features of the retrieved audio and T5-base features of their corresponding captions in an interleaving manner and feed into the encoder. With this retrieval setting, the TTA-RAG model moderately outperforms our TTA baseline on IS, CLAP and FAD scores (row 5 vs. row 4). Note that our Audiobox TTA baseline is significantly stronger than the AudioLDM TTA baseline, and the performance gap is not as large as in Re-AudioLDM (row 2 vs. row 1). 
However, Re-AudioLDM retrieves \emph{labeled} audio and requires both audio and their captions as conditional inputs during generation. On the other hand, our proposed TTA-RAG approach can retrieve from \emph{unlabeled} audio databases. In addition to the input text, only retrieved audio samples are included as conditional inputs. Our approach does not require the corresponding captions of the retrieved audio as the conditional inputs.

In Section \ref{sec:res-zero}, we have seen that zero-shot and few-shot TTA performance can be significantly improved using retrieved audio. As an in-domain setting, we evaluate the best Audiobox TTA-RAG model (i.e. trained with A2A - AS retrieval) on AudioCaps test set, with different retrieval configurations at inference time (row 6-11). We find that in general there is \emph{a trade-off between CLAP score and KL and FAD}: T2A retrieval leads to a significant CLAP score boost, at the cost of higher KL and FAD. We argue that a higher KL and FD does not necessarily mean bad audio quality. First, the high IS score suggests that the output audio is not noisy or silent. Second, KL and FAD measure the distribution distance between the target and generated audio. With retrieved audio as additional inputs, the output distribution can be shifted toward the retrieved audio and away from the target audio. One piece of evidence is that T2A - AC retrieval, which retrieves from the closer distribution of AudioCaps, leads to lower KL and FAD than T2A - AS retrieval. This suggests that the quality of retrieved samples is an important contributing factor to the output quality. Nevertheless, we argue that our proposed TTA-RAG model generates audio that semantically aligns with the input text, as shown by the high CLAP scores.

\section{Conclusion}
We have presented Audiobox TTA-RAG, a novel retrieval-augmented TTA approach. 
We showed that trained with the additional acoustic conditioning contexts from the A2A retrieval samples, Audiobox TTA-RAG significantly outperforms a strong TTA baseline in zero-shot and few-shot settings. 
For the in-domain setting, the proposed model can generate audio that is highly semantically aligned to the input text, as suggested by the significant improvement on CLAP scores. 
Our proposed method can retrieve from unlabeled audio data sources, offering more flexibility in practice and enabling large-scale retrieval on in-the-wild and unlabeled audio datasets. Through our extensive experiment, we demonstrated that our proposed model can effectively utilize the retrieved audio samples and benefit practical use cases.

\bibliographystyle{IEEEtran}
\bibliography{mybib}

\end{document}